\documentclass[12pt]{article}
\usepackage{amssymb}
\usepackage{amsmath}
\usepackage{graphicx}
\usepackage{subfigure}

\def\beq{\begin{eqnarray}}    
\def\eeq{\end{eqnarray}}      
\def\nn{\nonumber}    
\DeclareMathOperator{\cx}{\square}

\def\al{\alpha}
\def\be{\beta}

\def\de{\delta}

\def\La{\Lambda}
\def\la{\lambda}
\def\na{\nabla}
\def\pa{\partial}

\def\si{\sigma}

\def\ph{\varphi}

\def\Ga{\Gamma}
\def\De{\Delta}
\def\La{\Lambda}

\begin{document}

\begin{center}

{\large\sc
Bounce and stability in the early cosmology with anomaly-induced
corrections}
\vskip 6mm

{\bf Wagno Cesar e Silva}
\footnote{E-mail address: \ wagnorion@gmail.com}
\quad
and
\quad
{\bf Ilya L. Shapiro}
\footnote{On leave from Tomsk State Pedagogical University.
E-mail address: \ ilyashapiro2003@ufjf.br}
\vskip 6mm

{\small\sl Departamento de F\'{\i}sica -- ICE,
Universidade Federal de Juiz de Fora
\\
36036-100, Juiz de Fora, MG, Brazil}

\end{center}

\vskip 2mm

\begin{quotation}
\noindent
{\large\it Abstract}.$\,\,$
An extremely fast exponential expansion of the Universe is typical
for the stable version of the inflationary model, based on the
anomaly-induced action of gravity. The total amount of exponential
$e$-folds could be very large, before the transition to the unstable
version and the beginning of the Starobinsky inflation. Thus,
the stable exponential expansion can be seen as a pre-inflationary
semiclassical cosmological solution. We explore whether this
stable phase could follow after the bounce, subsequent to the
contraction of the Universe. Extending the previous consideration
of the bounce, we explore both stable expansion and the bounce
solutions in the models with non-zero cosmological constant and
the presence of background radiation. The critical part of the
analysis concerns stability for small perturbations of the Hubble
parameter. It is shown that the stability is possible for the
variations in the bounce region, but  not in the sufficiently
distant past in the contraction phase.
\end{quotation}
\vskip 4mm

\section{Introduction}

The importance of singularities in general relativity is partially based
on the fact that they are unavoidable \cite{Penrose-sing,Hawking-sing}
and, therefore, indicate to the limits of applicability of the theory. The
common belief is that, by properly extending general relativity, one
arrives at the singularity-free theory. In particular, the cosmological
singularities (see, e.g., \cite{Guth-Vilenkin-2003} and references
therein) led to various theoretical
developments, such as, e.g., exploration of $f(R)$-type models with
simplified form of quantum corrections \cite{GurStar79,MullerStar},
string cosmology \cite{Gasperini2002},
different versions of cyclic universe (see, e.g., \cite{Steinhardt2001}),
quantum cosmology (see, e.g., \cite{Nelson-WFotU-98,Casadio-98})
and, in general, the interest in the bouncing models
(see \cite{Novello2008,Peter} for reviews and many references).
Typically, bouncing models require fundamentally new assumptions,
such as the specially designed non-localities (see, e.g.,
\cite{Biswas2012,Koshelev2013}) or introducing a scalar field with
specially adjusted features (or a modified $f(R)$-type gravity, see,
e.g., \cite{Zhuk}). One of the most economical ways to avoid
singularity is the assume a relevant positive curvature of the space
section of the spacetime manifold \cite{Ellis2002} in the very early
Universe.

The natural question is whether one can achieve the
cosmological solutions with bounce with the minimal set of, or
even without any, additional assumptions. The proposal of this
type was done in the paper \cite{anju}. In the present paper we
elaborate the same idea in more detail and consider it from the
modern perspective of the effective field theory framework.

Even without quantum gravity, the consistent treatment near the
singularity requires taking into account the back reaction of the
quantum matter fields on the classical gravitational background.
In general, the derivation of the semiclassical corrections to the
gravitational action is an unsolved problem, except for an extreme
UV, where at least the one-loop effects are controlled by the
logarithmic divergences, admitting a simple and elegant
description in terms of conformal anomaly \cite{duff,duff94}
and anomaly-induced effective action \cite{rie,frts84}. It was
suggested that these quantum corrections may be sufficient to provide
a bouncing solution \cite{anju}, without additional assumptions.
In the present work, we elaborate this proposal in detail. First, we
extend the information about the bounce solution by including
the large cosmological constant and the background radiation. These
aspects of the model are relevant because a huge cosmological
constant may be generated in the UV by symmetry restoration, while
an intensive creation of matter from the vacuum is expected in the
region of bounce, when the Universe leaves the de Sitter-type
contraction phase. The second important point is the generalization
of the phase diagrams for unstable \cite{star} and stable
\cite{OPC-asta,StabInstab} phases, in the presence of the
cosmological constant. Finally, we explore the stability of the
approximately exponential expansion and contraction with respect
to the small perturbations of the conformal factor of the metric.
Taken together, these new results extend our general understanding
of the possibility of the bounce due to the semiclassical
contributions to the gravitational action.

The paper is organized as follows.  In sec.~\ref{s2}, we briefly
review the anomaly and induced action, and describe the derivation
of the $00$ component of the gravitational equations in the presence
of radiation from the conservation law. In sec.~\ref{s3}, we derive
the particular de Sitter-like solutions from the $00$ equations with
semiclassical correction and formulates the detailed conditions of
their stability. These conditions can be applied to both expanding
and contracting phases. In sec.~\ref{s4}, we present the phase
diagrams with the cosmological constant for stable and unstable
versions of the theory, and also give new examples of the bounce,
including with the background radiation or cosmological constant.
Sec.~\ref{s5} is devoted to the stability analysis of the contracting
case, clarifying the status of the bounce solutions. Finally, in
sec.~\ref{s6}, we draw our conclusions and also discuss general
perspectives of solving the problem of cosmological singularities
using quantum corrections.

The definitions of geometric quantities, used below, included the
signature $(+,-,-, -)$ of the metric, the curvature tensor
$R^\la_{\,\,\,\tau\al\be} = \pa_\al\,\Ga^\la_{\,\,\tau\beta} + \,...$,
and the scalar curvature $R = g^{\al\be}R^\la_{\,\,\,\al\la\be}$.

\section{Anomaly-induced action and the 
early Universe}
\label{s2}

In the very early epoch of the Universe, the typical energy scale
is very high. Different models of particle physics give different
estimates for the upper bounds of the masses of the particles,
but usually these upper bounds are of the order of magnitude, or
below, the GUT scale $M_X \propto 10^{16} GeV$, that is
a few orders below the Planck scale $M_P$. The Friedmann
equation, based on general relativity, predicts an unbounded
growth of the Hubble parameter near the beginning of time $t=0$,
and the singularity emerges with $H \sim t^{-1}$. Thus, in the
region close to the spacetime singularity, the gravitational field is
so intensive, that all real particles can be regarded massless. The
same concerns the virtual particles, which are interacting with the
strong gravitational background. Thus, the relevant quantum
effects in the vicinity of cosmological singularity may be limited
by the effects of  massless fields, which is, in fact, the simplest
case.

It is a natural hypothesis that the fundamental theory of particle
physics possesses the asymptotic freedom in the UV, including in
the vicinity of the singularity. Furthermore, for fermion and vector
fields, the masslessness implies local conformal invariance. This
is a natural feature from the perspective of quantum theory. The
point is that, free massless particles have a vanishing trace of the
energy-momentum tensor. Demanding the field-particles duality
in the description of quantum matter, we arrive at the condition
for the trace of the dynamical energy-momentum tensor,
\beq
T^\mu_{\,\mu}
\,=\,
-\,\frac{2}{\sqrt{-g}}\,g_{\mu\nu}\,\frac{\de S}{\de g_{\mu\nu}}
\,=\,0.
\label{Tmn-trace}
\eeq
On the other hand, (\ref{Tmn-trace}) is the Noether identity for
the local conformal symmetry
\beq
g_{\mu\nu}= e^{2\si} {\bar g}_{\mu\nu},
\qquad
\Phi_i =  e^{k_i \si} {\bar \Phi}_i,
\qquad
\si = \si(x),
\label{Noether-conf}
\eeq
where $k_i = (-1, \,-\frac32,\,0)$ is the conformal weight
for the matter fields (scalars, fermions and vectors,
$\Phi_i=(\ph,\,\psi,\,A_\mu)$, correspondingly. If the Noether
identity (\ref{Tmn-trace}) is satisfied, the action $S$ does not
depend on the conformal factor $\si(x)$. For the scalar
field, the symmetry (\ref{Noether-conf}) imposes an additional
condition on the value of the nonminimal parameter $\xi = \frac16$
of interaction with the scalar curvature, such that the Lagrangian of
this field (we write down the Lagrangian for a single real scalar, for
simplicity), becomes
\beq
{\cal L}_{scal}
\,=\,
\frac12\, g^{\mu\nu} \pa_\mu\ph \pa_\nu\ph
\,+\,\frac12\,\xi R\ph^2.
\label{Lscal-conf}
\eeq

Furthermore, it is sufficient to choose the action of vacuum to
satisfy the local conformal symmetry (\ref{Noether-conf}), without
jeopardizing the renormalizability of the theory. Taking the power
counting arguments into account, this means that the part of the
vacuum action, which is the subject of an infinite renormalization
in the UV, has the form
\beq
S_{cv}\, =\, \int d^4x\sqrt{-g}\,
\left\{ a_1 C^2 + a_2 E_4 + a_3 \cx R \right\},
\label{higher}
\eeq
where the integrand of the Gauss-Bonnet topological term and the
square of the Weyl tensor are
\beq
&& E_4=R_{\al\be\mu\nu}R^{\al\be\mu\nu}
-4 R_{\al\be}R^{\al\be} + R^2,
\nn
\\
&& C^2=R_{\al\be\mu\nu}R^{\al\be\mu\nu}
- 2 R_{\al\be}R^{\al\be} + \frac13 R^2,
\nn
\eeq
respectively. Action (\ref{higher}) satisfies the Noether identity
(\ref{Noether-conf}) at the classical level. On the top of this, the
classical action of vacuum may include the non-conformal part,
e.g., the Einstein-Hilbert term with the cosmological constant
$\La$, and the $R^2$ term,
\beq
S_{ncv}\,=\,
\int d^4x\sqrt{-g}\,\Big\{
-\,\frac{1}{16\pi G}\,(R + 2\La) + a_4R^2\Big\}.
\label{Einstein}
\eeq
Finally, $\,a_1,\,a_2,\,a_3,\,a_4$, $\,G\,$ and $\,\La\,$ are the
parameters of the vacuum action. The renormalization of the
parameters $\,a_1,\,a_2$ and $a_3$ generates the anomalous
violation of the Noether identity (\ref{Noether-conf}). The main
idea of the corresponding inflationary model is to use the anomaly,
or the anomaly-induced action of gravity, to arrive at the
cosmological solution which replaces the usual
Friedmann-Lema\^{\i}tre solution of general relativity.

Consider the particle physics theory with $N_0$ real scalars
(a complex scalar counts as two real scalars), $N_{1/2}$ Dirac
spinors, and $N_1$ massless vector fields. The beta functions
for the parameters $\,a_1,a_2,a_3$ are as follows:
\beq
&&
w \,=\, \be_1 \,=\, \frac{1}{(4\pi)^2}\,\Big(
\frac{N_0}{120} + \frac{N_{1/2}}{20} + \frac{N_1}{10} \Big)\,,
\label{w}
\\
&&
b \,=\, \be_2 \,=\, -\,\frac{1}{(4\pi)^2}\,\Big( \frac{N_0}{360}
+ \frac{11\,N_{1/2}}{360} + \frac{31\,N_1}{180}\Big)\,,
\label{b}
\\
&&
c \,=\, \be_3
\,=\, \frac{1}{(4\pi)^2}\,\Big( \frac{N_0}{180} + \frac{N_{1/2}}{30}
- \frac{N_1}{10}\Big) ,
\label{c}
\eeq
where we introduced the useful notations for the coefficients
$\,w$, $\,b\,$ and $\,c$. At the quantum level, for free matter fields,
the Noether identity  (\ref{Noether-conf}) is violated by the anomaly
in the vacuum sector,
\beq
T\,\,=\,\, \langle T_\mu^\mu \rangle \,\,=\,\,- \frac{2}{\sqrt{-g}}\,g_{\mu\nu}
\frac{\de {\bar \Ga}_{ind}}{\de g_{\mu\nu}}\,\,=\,\,
- \,\big(wC^2 + bE_4 + c \cx R + \be F^2 \big)\,.
\label{main equation}
\eeq
Here ${\bar \Ga}_{ind}$ is the renormalized quantum correction
to the classical action (\ref{higher}) plus the contribution to the
radiation term,  $F^2=F_{\mu\nu}F^{\mu\nu}$. One has to include
the background  radiation term because in the early Universe there
may be radiation. The radiation term is conformal invariant and
can be kept together with $C^2$ term in the derivation of
anomaly-effective action ${\bar \Ga}_{ind}$. The general form
of the beta function is
\beq
\beta
&=&
-\,\dfrac{2e^{2}}{3(4\pi)^{2}}N_{f}-\dfrac{e^{2}}{6(4\pi)^{2}}N_{s},
\label{beta}
\eeq
where $N_{f}$ and $N_{s}$ are multiplicities of charged fermions and
scalars.

Solving Eq.~(\ref{main equation}) with respect to ${\bar \Ga}_{ind}$
is a relatively simple exercise, described in many papers (e.g., in
the first original works \cite{rie,frts84}), the review \cite{PoImpo},
book \cite{book} and the forthcoming textbook \cite{QFT-OUP}).
Here we need just the simplest version of the solution, which can
be obtained directly from the two relations
\beq
&& \frac{\de A}{\de \si(x)}=
\int d^4 y \,\,\frac{\de g_{\mu\nu}(y)}{\de \si(x)}\,
\frac{\de A}{\de g_{\mu\nu}(y)}
\,\,=\,\,
2\,g_{\mu\nu}\,\frac{\de A}{\de g_{\mu\nu}}
\nonumber
\eeq
and
\beq
\sqrt{-g}\,\Big(E_4-\frac23\,{\cx} R\Big)
\,=\,
\sqrt{-g}\,\Big({\bar E_4}-\frac23\,{\bar \cx} {\bar R}
+ 4{\bar \De_4}\si \Big).
\label{119}
\eeq
In these relations, we use the conformal parametrization of the
metric (\ref{Noether-conf}).
$\De_4$ is the fourth-order, Hermitian, conformal invariant
operator \cite{FrTs-superconf,Paneitz} acting on a
conformal invariant, dimensionless scalar field,
\beq
\De_4 = \cx^2 + 2\,R^{\mu\nu}\na_\mu\na_\nu - \frac23\,R{\cx}
+ \frac13\,(\na^\mu R)\na_\mu.
\label{120}
\eeq

The general solution for the anomaly-induced effective action is
\footnote{In this paper, we deal only with the dynamics of the
conformal factor. Thus, this simplest non-covariant solution is
equivalent to the nonlocal covariant one \cite{rie,frts84}
and to the local covariant with two auxiliary fields \cite{a}.}
\cite{rie,frts84}
\beq
&& \Ga_{ind}=S_c[{\bar g}_{\mu\nu}] \,+\,
\int d^4 x \sqrt{-{\bar g}}\,\Big\{
\,\si \big( w{\bar C}^2 + \be {\bar F}^2  \big)
+ b\si \Big({\bar E}-\frac23 {\bar \cx}
{\bar R}\Big) + 2b\si{\bar \De}_4\si\Big\}
\nn
\\
&& \qquad -\,
\frac{3c+2 b}{36}\,\int d^4 x\sqrt{-g}\,R^2,
\label{quantum1}
\eeq
where $\sqrt{-g} = \sqrt{-{\bar g}} e^{4\si}$ and
$R = e^{-2\si}\big[{\bar R}
- 6({\bar \na}\si)^2 - 6 \bar {\cx} \si \big]$.
The first term in the solution (\ref{quantum1}) is an arbitrary
conformal invariant functional,
$\,S_c[{\bar g}_{\mu\nu}]=S_c[g_{\mu\nu}]$, which plays the role
of an integration constant for the equation (\ref{main equation}).
There is no regular method for deriving $S_c[{\bar g}_{\mu\nu}]$
in the explicit form. However, in the known physical applications it
is usually considered insignificant. E.g., for the background
cosmological solutions (not for perturbations), this functional is
irrelevant because an isotropic and homogeneous metric has the form
\beq
g_{\mu\nu}={\bar g}_{\mu\nu}\cdot a^2(\eta).
\label{flat metric}
\eeq
As usual, the conformal time $\eta$ is related to the cosmic time $t$
by the formula $\,a(\eta)d\eta=dt$. Thus, $S_c[{\bar g}_{\mu\nu}]$
is obviously independent of the conformal factor $a(\eta)$.

The form of Eqs.~(\ref{main equation}) and (\ref{quantum1})
demonstrates that the $\cx R$ term in the anomaly contributes
to the covariant local $R^2$ term in the induced action. The
coefficient of this term is regularization-dependent
\cite{birdav,duff94}. This dependence was explored in detail
in \cite{anomaly-2004}, where it was shown, in particular, that
it is equivalent to modifying the finite $a_4$ term in the classical
non-conformal action (\ref{Einstein}).

Different from the $C^2$ term in the action (\ref{higher}), the
$R^2$ term affects the equation for the conformal factor $\,a(\eta)$.
Therefore, it is important to fix the value of $a_4$ by some
physical conditions. In the literature, there are two main choices
for such a condition. The most important one is
$a_4 \approx 5 \times 10^8$, providing the consistent inflationary
model of Starobinsky \cite{star,star83}. The large magnitude of
this coefficient, compared to the beta function (\ref{c}), makes the
quantum contributions to the $R^2$ term completely irrelevant.
Another interesting choice is $a_4 \approx 0$, such that the
particle contents $N_0$, $N_{1/2}$, $N_1$ in (\ref{c}) becomes
an important issue. In what follows, we shall follow this option
and discuss how the choice of the particle contents defines the
stability of the cosmological solutions.

The basis of the cosmological model for the early Universe is the
total action, including  (\ref{Einstein}), (\ref{higher}), matter and
the quantum corrections described by (\ref{quantum1})
\beq
S_t=S_{vacuum}+S_{matter}+\Ga_{ind}.
\label{massless}
\eeq
Let us remember that the matter corresponds to the pure radiation,
its action is conformal invariant and thus, it does not affect the
dynamical equation\footnote{We shall see how the radiation density
appears in the first integral of this equation.} for $\,a(\eta)$.

Taking variational derivative with respect to
$a(\eta)$, we arrive at the equation of motion in the form
\beq
T_{EH}\,+\,T_{HD}\,=\,T_{radiation},
\label{T-eq}
\eeq
where the traces of the equations of motion for the corresponding
terms in the action, are
\beq
&&
T_{EH}
\,=\,
\frac{3}{4\pi G} \Big(\frac{a^{\prime\prime}}{a^3}
+ \frac{k}{a^{2}} -\frac{2 \Lambda}{3}\Big)\,,
\label{TEH}
\\
&&
{\bar T}_{HD}
\,=\,
6c \left[
\frac{4a^{\prime\prime\prime}a^{\prime}}{a^6}
-\frac{a^{(4)}}{a^5}
+ \frac{3{a^{\prime\prime}}^2}{a^6}
- \frac{6a^{\prime\prime} a^{\prime\, 2}}{a^7}
+ 2k\Big(\frac{a^{\prime\prime}}{a^{5}}
- \frac{{a^{\prime}}^{2}}{a^{6}}\Big) \right]
\nonumber
\\
&&
\qquad
\,\,+\,24b\left[\frac{a^{\prime\prime} a^{\prime \, 2}}{a^7}
- \frac{{a^\prime}^4}{a^8}
+ k\Big(\frac{a^{\prime\prime}}{a^{5}}
- \frac{{a^{\prime}}^{2}}{a^{6}}\Big)\right],
\label{THD}
\\
&&
T_{radiation}
\,=\, \frac{\beta {\bar F^2} }{a^4}\,,
\label{Trad}
\eeq
where $k=0$ or $k=\pm 1$ for different space geometries.

Eq.~(\ref{T-eq}) represents the trace of the equations of motion, and
can be used for analytical and numerical study of the system. However,
it is more useful to derive $00$-component $\rho_i$ using the
conservation law for each of the components $T_{EH}$, $T_{HD}$
and $T_{radiation}$,
\beq
d\,(\rho_i \,a^3) =-p_i \,d(a^3) ,
\quad \mbox{where} \quad
T_{i} = \rho_i - 3 p_i.
\label{conslaw}
\eeq

In what follows, we shall use the notations and general approach of
the previous works, e.g., \cite{fhh,RadiAna}. Let us stress that this
consideration implies that we omit the conformal invariant functional
$\,S_c[g_{\mu\nu}]$ in the purely gravitational part of the
expression (\ref{quantum1}). This approximation is necessary here,
because all the information we have, comes from the anomaly and
reflects only the trace of the equations of motion.  From
Eq.~(\ref{conslaw}), follows that
\beq
\frac{d\rho_i}{da^3}+\frac{4}{3}\frac{\rho_i}{a^3}=\frac{T_i}{3a^3},
\label{density equation}
\eeq
with the general solution defined by
\beq
\rho_i(a)=C(a)a^{-4},
\quad \mbox{with} \quad
\frac{d C}{d\eta} = T_i \,a^3\,a^{\prime}\,.
\label{coef. equation}
\eeq

Integrating (\ref{coef. equation}) for each of the components,
we arrive at the following results:
\beq
&&
\rho_{EH}
\,=\,
\frac{3}{8\pi G}\,\Big(\frac{{a^\prime}^2}{a^4} + \frac{k}{a^{2}}
- \frac{\Lambda}{3}\Big)
\,=\,
\frac{3M_P^2}{8\pi}\,\Big(\frac{{\dot a}^2}{a^2} + \frac{k}{a^{2}}
- \frac{\Lambda}{3}\Big)
\nn
\\
&&
\qquad
\,=\,
\frac{3M_P^2}{8\pi}\,\Big( {\dot \si}^2 + ke^{-2\si} - \frac{\Lambda}{3}\Big),
\label{EOS EH}
\\
&&
{\bar \rho}_{HD}
= 
\frac{6b \big({a^{\prime}}^4 + 2ka^{2} {a^{\prime}}^{2} \big)}{a^8}
+ \frac{3c\big(a{a^{\prime\prime}}^2
	-  2a a^{\prime\prime\prime}a^{\prime}
	+ 4 a^{\prime\prime}{a^\prime}^2 + 2k a {a^{\prime}}^{2}\big)}{a^7}
\nn
\\
&&
\qquad
\,=\,
(6b+9c)\frac{{\dot a}^4}{a^4}
+ 6k(b+c)\frac{\dot a^{2}}{a^{4}} + 3c\Big(
\frac{{\stackrel{..}{a}}^{2}}{a^{2}}
- \frac{2 \stackrel{..}{a}{\stackrel{.}{a}}^2}{a^{2}}
- \frac{2 \stackrel{...}{a}\stackrel{.}{a}}{a^{2}}
\Big)
\nn
\\
&&
\qquad
\,=\,
6b {\dot \si}^4
+ 6k(b+c)\dot \si^{2}e^{-2\si} + 3c\Big({\stackrel{..}{\si}}^2
- 2 \stackrel{...}{\si} \stackrel{.}{\si}
- 6 \stackrel{..}{\si}{ \stackrel{.}{\si} }^2
\Big),
\label{EOS HD}
\\
&&
\rho_{radiation}
\,=\,
\frac{\rho_{r0}+\be{\bar F}^2\ln a}{a^{4}}
\,=\,
(\rho_{r0}+\be{\bar F}^2\si) e^{-4\si}.
\label{EOS R}
\eeq
In these expressions we have changed from conformal time $\eta$ to the
cosmological time $t$, from $a(t)$ to  $\si(t)=\ln a(t)$ and used the
definition of Planck mass $\,M_P^2=1/G$. One of the advantages of
the $00$-component of the equations of motion compared to the trace
is that, in the expression (\ref{EOS R}), we can include the classical
radiation term. Since the corresponding term is conformal, it was
hidden in the trace of the equations (\ref{T-eq}).

\section{De Sitter-like solutions and their stability}
\label{s3}

Now we are in the position to write the dynamical equation for
$\si(t)$ in the form
\beq
\rho_{EH}\,+\,\rho_{HD}\,=\,\rho_{radiation},
\label{rho-eq}
\eeq
where the $\rho_i$ are defined in Eqs.~(\ref{EOS HD}), (\ref{EOS EH})
and (\ref{EOS R}). Most of the further discussion will be based on
this equation. For the sake of brevity, we restrict our attention to
the $k=0$ case.

It is most economical
to write the purely gravitational part of the equation in terms of the
Hubble parameter $H={\dot \si}$. In this way we get
\beq
\frac{3M_P^2}{8\pi}\,H^2 - \frac{\La M_P^2}{8\pi}
+6bH^4 + 3c \big(  {\dot H}^2 - 2 H{\ddot H} - 6 H^2{\dot H} \big)
= (\rho_{r0}+\be{\bar F}^2\si) e^{-4\si}.
\label{H-eq}
\eeq
As the first step, consider the equation without the radiation term.
Looking for the solution with a constant $H$, we meet a biquadratic
equation
\beq
6bH^4 + \frac{3M_P^2}{8\pi}\,H^2 - \frac{\La M_P^2}{8\pi}
\,=\, 0,
\label{H0-eq}
\eeq
with the following four solutions \cite{asta}:
\beq
H\,=\, \pm\,\frac{M_P}{\sqrt{-32\pi b}} \, \bigg(\,1\pm \,
\sqrt{1+\frac{64\pi b}{3}\,\frac{\Lambda }{M_P^2}\,}
\bigg)^{1/2}\,.
\label{H}
\eeq
It is clear that the solutions with the positive $H$ correspond
to expanding, and the ones with the negative $H$, to contracting
Universes. From Eq.~(\ref{b}) follows that $b<0$
for any particle contents of the underlying theory, such that  the
$\sqrt{-32\pi b}$ is always real. Consider the sign in the
parenthesis, taking $H>0$ and $\La>0$ for the definiteness. Taking
into account that $\La \ll M_P^2$, we get the two approximate
solutions:
\beq
H_1\,=\,\sqrt{\frac{\Lambda }{3}}
\qquad
{\rm and}
\qquad
H_2\,=\, \frac{M_P}{\sqrt{- 16\pi b}}.
\label{HH}
\eeq
The first one, $H_1$, is the classical de Sitter solution where the
anomaly-induced terms play no role. The second solution,  $H_2$,
is the solution corresponding to the equilibrium between quantum,
anomaly-induced part and the classical part, of Eq.~(\ref{H-eq}).
This solution can not be obtained in the approach when the quantum
part is treated a small perturbation. On the other hand, it can be
derived from different approaches to quantum corrections in
curved spacetime \cite{DoCr,FrTs84}.

Let us consider the stability of these solutions. Our purpose is to
explore the asymptotic stability, and, owing to the exponential time
dependence of $a(t)$, it is important to perform variation of $\si(t)$
or $H(t)$, instead of $a(t)$ \cite{asta}. For the variation of
$H(t) \to H(t) + X(t)$, without the radiation term in (\ref{H-eq})
and in the linear approximation, we obtain
\beq
24bH^3X + 6c{\dot H}{\dot X} - 6c{\ddot H}X - 6c{\ddot X}H
- 18c{\dot X}H^2 - 36c{\dot H}HX\,+\,\frac{3M^2_P}{4\pi}HX
\,=\,0.
\quad
\label{pert H full}
\eeq
For a constant $H=H_0$, such as the ones in (\ref{HH}), this
equation boils down to
\beq
{\ddot X} + 3H_0{\dot X} - \Big(\frac{4bH_0^2}{c}
+ \frac{M_P^2}{8\pi c}\Big)X\,=\,0.
\quad
\label{pert H}
\eeq

In the case of extreme inflationary solution with $H_0=H_2$ in
(\ref{HH}), we have $-\frac{4bH_0^2}{c}= \frac{M_P^2}{4\pi c}$
and then the solution of (\ref{pert H}) has the form
\beq
X\,=\,C_1 e^{\la_1t}\,+\,C_2 e^{\la_2t},
\quad
\label{solapm}
\eeq
where
\beq
\la_{1/2}
\,=\,-\frac{3H_0}{2} \pm \frac{3H_0}{2}\sqrt{1+\frac{8b}{9c}}.
\label{lapm}
\eeq
Since we always have $b<0$, for $c>0$ there are no eigenvalues
with the positive real part hence there are no growing modes and
the solution with $H_0=H_2$ is stable \cite{star,anju}. Using the
equation for the trace, one can show that in the model with $c<0$
the same solution is unstable.

It is instructive to consider the stability of the classical solution
with $H_0=H_1$. In this case the equation for linear perturbation
has the form
\beq
{\ddot X} + 3H_0{\dot X} - \frac{M_P^2}{8\pi c}\,X\,=\,0
\quad
\label{pert H1}
\eeq
and the eigenvalues in (\ref{solapm}) are
\beq
\la_{1/2}
\,=\,- \frac{\sqrt{3\La}}{2} \pm \frac{M_P}{\sqrt{8\pi c}}
\left( 1 + \frac{6\pi c\La}{M_P^2}\right)^{1/2}
\,\approx\,- \frac{\sqrt{3\La}}{2} \pm \frac{M_P}{\sqrt{8\pi c}}.
\label{eigH1}
\eeq
Obviously, for $c>0$ there is a Planck order positive eigenvalue
and hence a very fast growing mode. For $c<0$ there are
oscillations with the frequency of the Planck order of magnitude,
suppressed by a relatively slow damping. Qualitatively, the results
for the stability are the same as derived earlier from the trace
equation (\ref{T-eq}) in \cite{asta}.

The anomaly-induced action (\ref{quantum1}) includes the
semiclassical corrections to the gravitational action which are
certain, in the sense they do not depend on the way we intend
to quantize gravity, use string theory, etc. Thus, we have to
correctly interpret the stability  results, formulated above. The
Starobinsky model of inflation \cite{star,star83} is based on the
unstable version and requires specially chosen initial conditions.
As we have mentioned above, the phenomenological consistency
requires adding a classical $R^2$ term with
$a_4\approx 5\times 10^8$ in the action (\ref{Einstein}). On the
other hand, the stable version is free from the restrictions
on the initial conditions.

As we just discussed, without the classical $a_4$ term, the stability
is defined by the sign of $c$ and this sign depends on the particle
contents, according to Eq.~(\ref{c}). The stability of $H_1$ solution
is achieved for $N_1 > 18N_0 + 3N_{1/2}$. This inequality holds for
the minimal standard model of particle physics (MSM) with $N_1=12$,
$N_0=4$ and $N_{1/2}=24$, but does not hold for supersymmetric
extensions, such as MSSM, or for the ``doubled'' standard model with
$N_1=12$, $N_0=8$ and $N_{1/2}=48$. In what follows, we shall use
the last version for the numerical analysis, but one has to remember
that there is no real difference between the behavior of different
models with $c>0$. Let us note that the $c=0$ case, which occurs for
the $N=4$ supersymmetric Yang-Mills theory \cite{Henningson-1998},
gives an unstable solution with $H_2$.

It is clear that, in the late Universe, there should be either $c<0$,
or a sufficiently large classical $a_4$, as otherwise flat space
would be unstable.\footnote{As we already mentioned above, these
two possibilities are equivalent \cite{birdav,duff94,anomaly-2004}.}
Thus, using the advantage of the stable version at the beginning of
inflation, requires the scheme of transition between stable and
unstable particle contents. One of the possibilities for such a
transition is related to the low-energy decoupling of the heavy
degrees of freedom beyond the MSM, e.g., the s-particles in the case
of MSSM \cite{susy-key}. A possible explanation of this decoupling
is based on the slowing down exponential expansion with
$a(t) \sim \exp\{H_2t\}$ owing to the masses of the quantum
s-particles \cite{Shocom,asta}. It turns out that the process of
slowing down inflation takes a long while and, as a result, the
observable effects of inflation come from the very last $60-70$
$e$-folds out of much greater rate of inflation.

The next question regarding this transition is how to provide
the sufficiently large coefficient $a_4$ at the physically relevant,
last 60-70, $e$-folds. This problem has been addressed in
\cite{StabInstab}
and will not be discussed here. Instead of this, in the rest of the
paper, we shall discuss the beginning of the exponential phase
and, in particular, whether it could follow after the bounce from
a contracting phase.

\section{Phase diagrams}
\label{s4}

Before starting the analysis of stability of the special solutions
with respect to the small perturbations, let us complete the
previous section by considering the phase diagrams for the
Eq.~(\ref{H-eq}). Without the radiation term the equation is
\beq
2H\ddot{H}+6H^{2}\dot{H}-\dot{H}^{2}
- \frac{2b}{c}H^{4}
- \frac{M_P^{2}}{8\pi c} \left(H^{2}
- \frac{\Lambda}{3}\right)=0.
\label{E1o}
\eeq
We shall follow the analysis of the original paper by Starobinsky
 \cite{star}, and also the ones in \cite{OPC-asta,StabInstab} for the
 stable case. The main difference with these publications is that
 we include the cosmological constant term. As before, we restrict
 the main attention to the $k=0$ flat-space case. Indeed, there are
no strong arguments of why the space curvature could be important
in the very early Universe.

On the other hand, the cosmological constant may be relevant in
the vicinity of the cosmological singularity because its value may
be dramatically different from the one in the late Universe. The
reason is that, in the last case, the observed value of the vacuum
energy density
$\rho_\La^{(obs)} = \frac{\La^{(obs)}}{16\pi G^{(obs)}}$
is a sum of the vacuum term $\rho_\La^{(vac)}$
and the contribution $\rho_\La^{(ind)}$, induced from the symmetry
breaking in the matter fields sector, e.g., from the spontaneous
breaking in the Higgs potential \cite{Weinberg89} (see also
\cite{CC-nova}).
In the early epoch, this symmetry can be (and most likely is)
restored owing to the high temperature or/and high curvature
effects. As a result, the   $\rho_\La^{(obs)}$ acquires the
value which is typical for the corresponding phase transition.
E.g., in case of the electroweak phase transition, this value
is about 56 orders of magnitude greater than the one in the
present Universe, but still negligibly small compared to the
Planck scale in the vicinity of the singularity. However, in
the case of the possible GUT-scale phase transition the difference
between   $\rho_\La^{(obs)} \propto M_X^4$ and $M_P^4$,
may be just a few orders, and in the case of  the Planck-scale
phase transition the two quantities may be of the same order
of magnitude. In what follows we consider the last two
possibilities.

\begin{figure}[h!]
\centering
\includegraphics[width=0.42\textwidth]{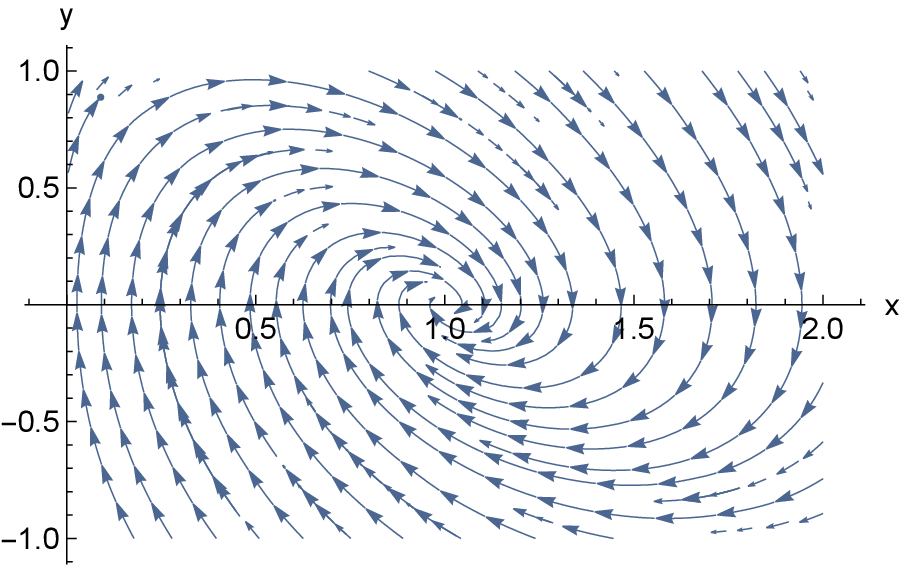}
\includegraphics[width=0.42\textwidth]{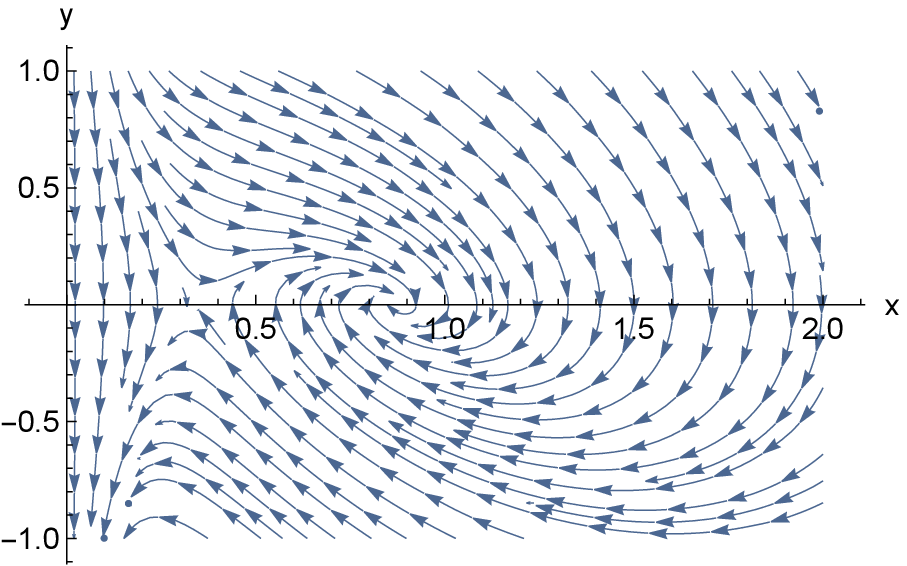}
\begin{quotation}
\caption{\small
Phase diagram for Eq.~\eqref{Eqf} with the MSSM
particle contents. In the left plot we choose $\La=0.001 M_P^2$ 
and in the right plot, $\La=0.5  M_P^2$.}
\label{df}
\end{quotation}
\end{figure}

\begin{figure}[h!]
\centering
\includegraphics[width=0.42\textwidth]{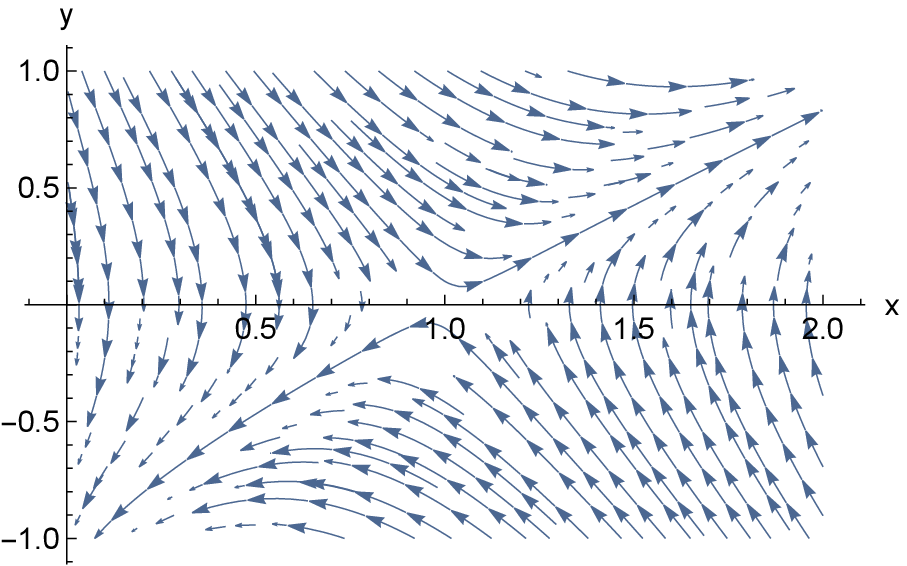}
\includegraphics[width=0.42\textwidth]{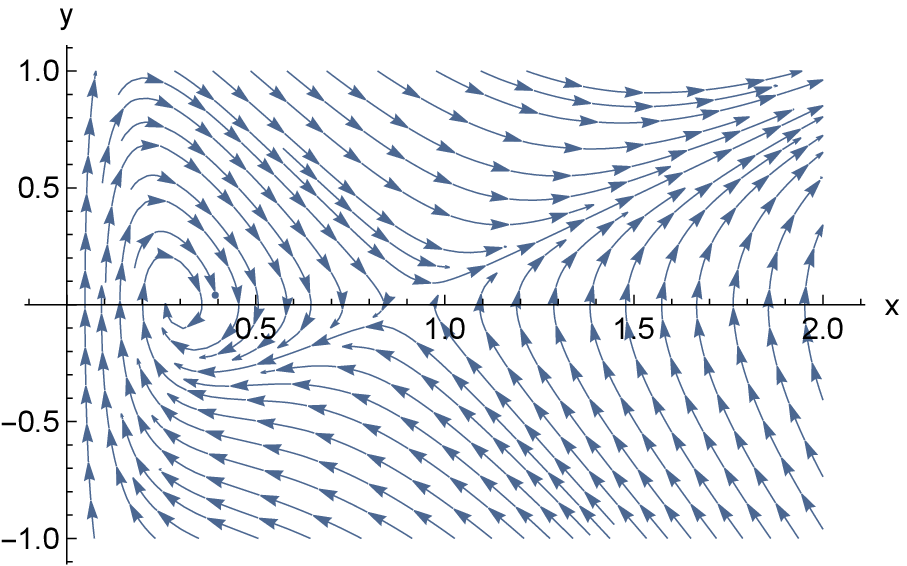}
\begin{quotation}
\caption{\small
Phase diagram for Eq.~\eqref{Eqf} with the MSM
particle contents, corresponding to the unstable exponential
solution. In the left plot there is $\La=0.001  M_P^2$
and in the right plot, $\La=0.5  M_P^2$.}
\label{dfinst}
\end{quotation}
\end{figure}

The order of Eq.~\eqref{E1o} can be reduced using the
following change of variables \cite{star}:
\beq
x=\left(\frac{H}{H_0}\right)^{\frac{3}{2}}
\hspace{0.35cm}
\textrm{and}
\hspace{0.35cm}
y=\frac{\dot{H}}{2H_{0}^{\frac{3}{2}}}H^{-\frac{1}{2}}.
\label{vS}
\eeq
Indeed, the reduction can be achieved by a simpler
transformation $\dot{H}=x(H)$, but the resulting equation
 is more complicated. Eq.~\eqref{E1o} can be elaborated
using   \eqref{vS} as
\beq
\dot{H}=\frac{2}{3}H_0x^{-\frac{1}{3}}\dot{x},
\qquad
\mbox{and}
\qquad
\ddot{H}=-\frac{2}{9}H_0x^{-\frac{4}{3}}\dot{x}^{2}
+\frac{2}{3}H_0x^{-\frac{1}{3}}\ddot{x}.
\label{MV1}
\eeq
In this way, we arrive at
\beq
\ddot{x}-\frac{2}{3}x^{-1}\dot{x}^{2}
+3H_0x^{\frac{2}{3}}\dot{x}
-\frac{3b}{2c}\,H^{2}_0x^{\frac{7}{3}}-\frac{3 M^{2}_{P}}{32\pi c} \left(x-\frac{\Lambda}{3}H^{-2}_0x^{-\frac{1}{3}}\right)=0.
\label{MV2}
\eeq
The second change of variables in \eqref{vS} gives
\beq
\dot{x}=3H_0x^{\frac{2}{3}}y
\quad
\mbox{and}
\quad
\ddot{x}
\,=\,
2H_0x^{-\frac{1}{3}}\dot{x}y+3H_0x^{\frac{2}{3}}\dot{y},
\label{MV3}
\eeq
such that the last relation becomes
\beq
\ddot{x}
\,=\,
6H^{2}_0x^{\frac{1}{3}}y^{2}+3H_0x^{\frac{2}{3}}
\frac{dx}{dt}\frac{dy}{dx}
\,=\,
6H^{2}_0x^{\frac{1}{3}}y^{2}+9H^{2}_0x^{\frac{4}{3}}
\,\frac{ydy}{dx}.
\label{MV4}
\eeq
Inserting these relations in \eqref{MV2}, we obtain
\beq
\frac{dy}{dx}
=
\frac{b}{6yc}\left[x+\frac{M^{2}_{P}H^{-2}_{0}}{16\pi b} \left(x^{-\frac{1}{3}}-\frac{\Lambda}{3}H^{-2}_0x^{-\frac{5}{3}}\right)\right]-1.
\label{Eqf0}
\eeq
Taking $ H_{0}=H_{+} $, defined in \eqref{H}, we can write
\beq
\frac{dy}{dx}
=
\frac{b}{6yc}\left[x-\frac{2}{\alpha}x^{-\frac{1}{3}}-\frac{64\pi b}{3\alpha^{2}M^{2}_{P}}
\Lambda x^{-\frac{5}{3}}\right]-1,
\label{Eqf}
\eeq
where we used an abbreviation 
$\al \equiv \left(1+\sqrt{1+\frac{64b}{3}\frac{\La}{M^{2}_{P}}}\right)$. 
The analogous formula from \cite{StabInstab} can be reproduced
in the limit $\,\La=0$. Using \eqref{Eqf}, we arrive to the phase
diagrams for the two different values of $\La$, as  shown in
Fig.~\ref{df} for the stable and in Fig.~\ref{dfinst} for the
unstable versions of the theory\footnote{All numerical calculations
and plots were done using Wolfram's Mathematica \cite{Wolfram}.}.

One can easily note that small cosmological constant does not
produce essential modifications compared to the previously known
$\La=0$ diagrams \cite{star} and \cite{OPC-asta}. At the same time,
the cosmological constant of the Planck order of magnitude, greatly
transforms the phase diagram of the stable case making it
qualitatively similar to the one of the unstable case with a small
cosmological constant. This feature suggests a new mechanism
of transition between the stable and unstable phases (different from
the one described in \cite{susy-key,Shocom} and \cite{StabInstab})
and perhaps deserves further investigation in the future.

\begin{figure}[h!]
\centering
\includegraphics[width=0.42\textwidth]{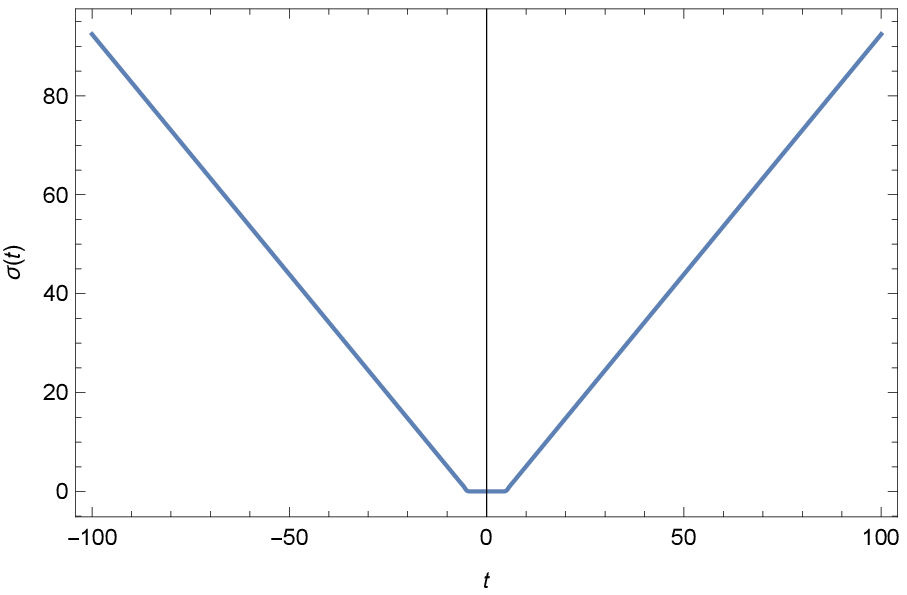}
\includegraphics[width=0.42\textwidth]{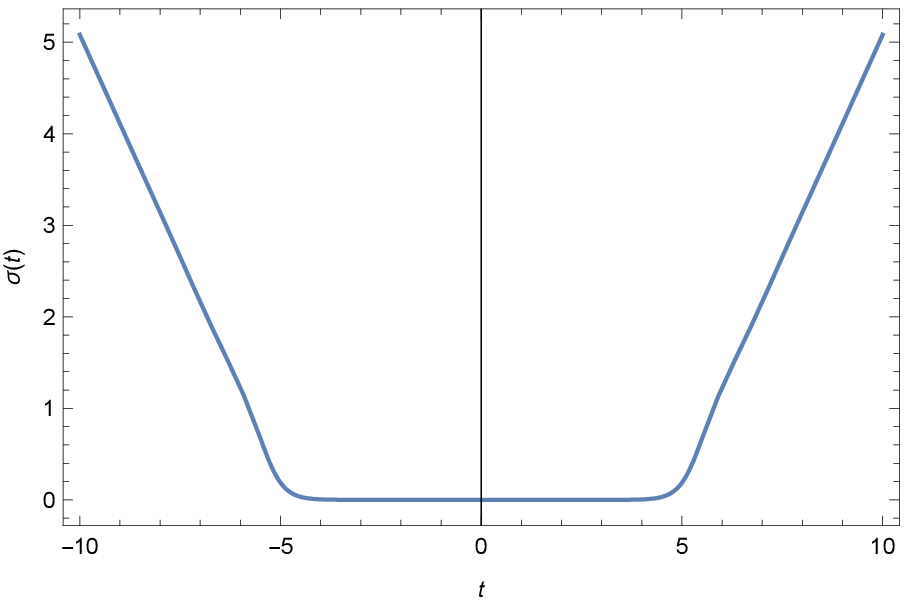}
\begin{quotation}
\caption{\small
Solution without radiation for the MSSM particles
contents and hence $c>0$. The initial conditions are
$ \sigma(0)=0$, $\dot{\sigma}(0)=-10^{-4}\,H_2$, $\ddot{\sigma}(0)=0$ and
$\dddot{\sigma}(0)=0$. On the left plot we show the interval
$-100 \leq t \leq  100$ in the Planck units and on the right plot
the interval is ten times smaller.}
\label{basic bc}
\end{quotation}
\end{figure}

Imposing the special initial conditions at $t=0$, one can arrive at
the bounce solutions, with the exponential [for a(t)] expansion
after the exponential contraction, and the smooth transition between
these two phases. The basic example of this sort is shown in
Fig.~\ref{basic bc} for the model with zero cosmological constant.
It is important that we cannot set
$\dot{\si}(0)=0$, as otherwise the solution is $\si(0)\equiv 0$.

\begin{figure}[h!]
\centering
\includegraphics[width=0.42\textwidth]{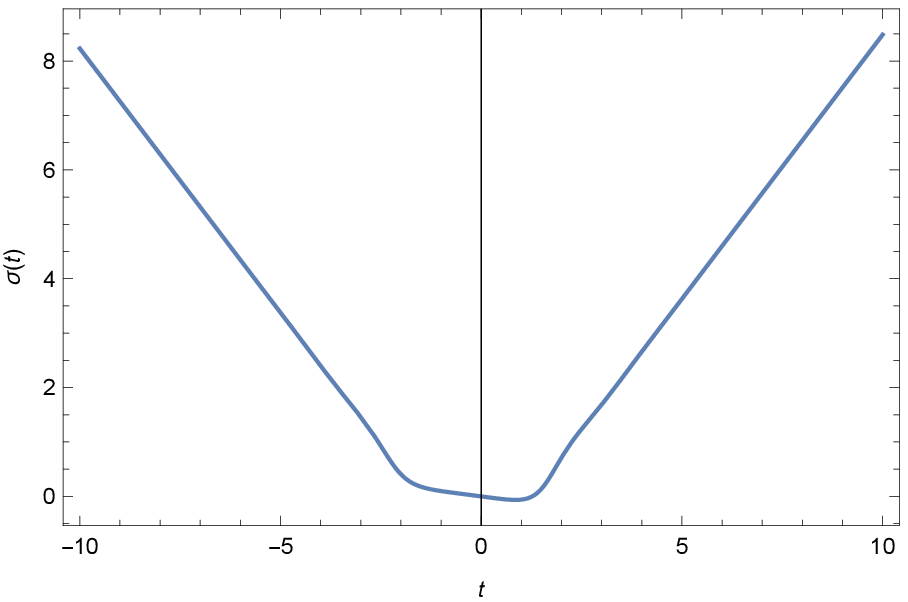}
\includegraphics[width=0.42\textwidth]{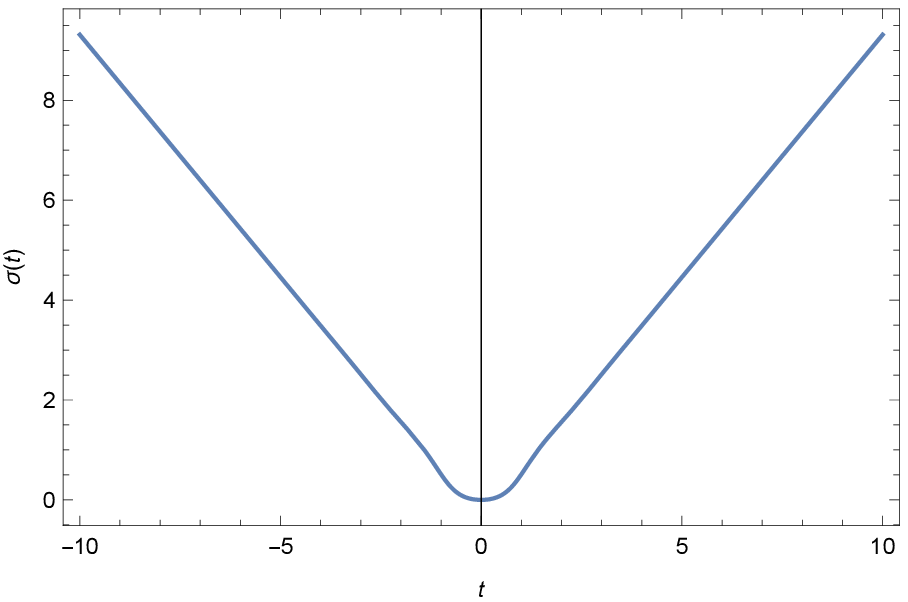}
\begin{quotation}
\caption{\small
Numerical solution for the conformal factor $\si(t)$ with the small
variation in the initial conditions at $t = 0$. The left plot
shows the behavior of the conformal factor under variations of
$\dot{\sigma}(0)$ and $\dddot{\sigma}(0)$. In this case, the initial
conditions become $\sigma (0) = 0$, $\dot{\sigma}(0) = -0.1H_ {2}$,
$\ddot{\sigma} (0) = 0$ and $\dddot{\sigma}(0) = 0.05$. In the
 right plot, the variation was performed as $\ddot{\sigma}(0)$, so
 that the new initial conditions are $\sigma (0) = 0$,
 $\dot{\sigma}(0) = -10^{-4}H_2$,  $\ddot{\sigma}(0) = 0.6$ and
 $\dddot{\sigma}(0) = 0$. In both cases the bounce solution is
 maintained.}
\label{b2}
\end{quotation}
\end{figure}

Let us present more details concerning the bounce solutions, based on
the numerical analysis. Fig.~\ref{b2} demonstrates the result of the
small variations of initial conditions. Here we still keep zero
cosmological constant. These plots clearly show that the bounce
survives such a small changes. However, we have found that larger
modifications may exclude the bounce solutions.

Another two relevant aspects concern the choice of initial
conditions. As we already know, changing these conditions, one can
produce the solutions without bounce. It is especially important,
that these conditions are imposed  in the vicinity of the minimum
$t=0$ of the curve. This point is in the past for the expansion
region $t>0$, but, at the same time, it  is in the future for the
region $t<0$ where we meet a contraction. In the next section, we
shall elaborate on the qualitative sense of this restriction.

\begin{figure}[h!]
\centering
{\label{brL1}
\includegraphics[width=0.42\textwidth]{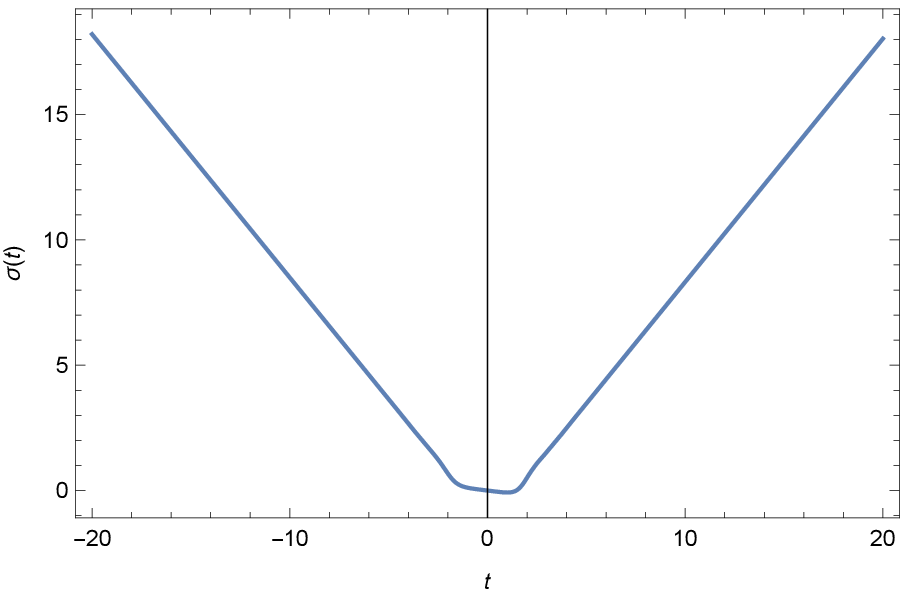}}
{\label{brL2} \includegraphics[width=0.42\textwidth]{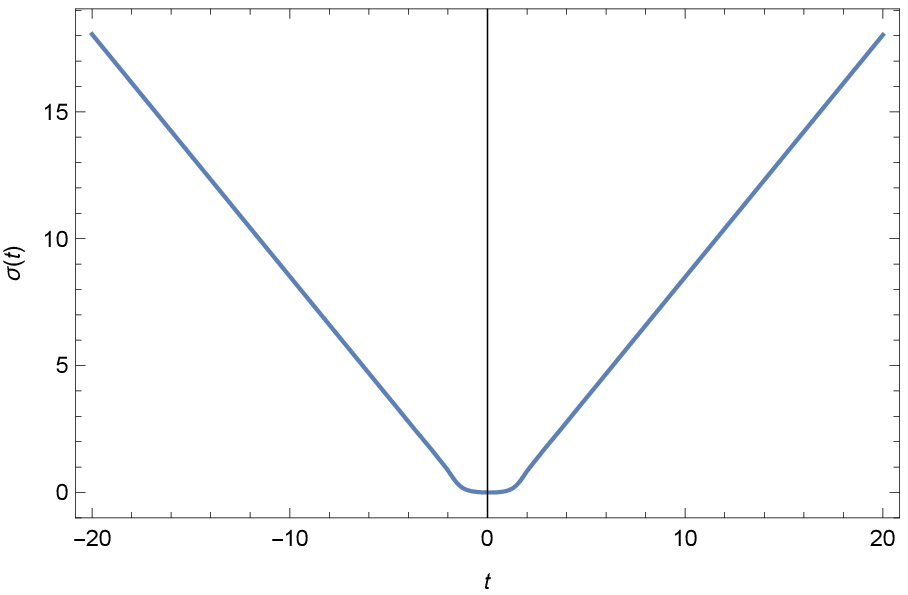}}
\begin{quotation}
\caption{\small
Numerical solution for the conformal factor $\si(t)$ with the non-zero
cosmological constant density. In the left graph, we consider
$\Lambda = 0.001 H^2_2$ and the initial conditions $\si(0) = 0$,
$\dot{\si}(0) = - 0.1H_2$, $\ddot{\si}(0) = 0$ and
$\dddot{\si}(0) = 0$. In the right plot, the value is larger,
$\La = 0.1 H ^2_2$ and the initial conditions are $\si(0) = 0$,
$\dot{\sigma}(0) = - 0.01H_ {2}$, $ \ddot {\sigma} (0) = 0.1$
and $\dddot {\sigma} (0) = 0$.}
\label{brL}
\end{quotation}
\end{figure}

The next step is to switch on the cosmological constant. As we have
mentioned above, in the region close to singularity the magnitude of
the corresponding density $\rho_\La^{(obs)}$ may be either a few
(even many, actually) orders of magnitude smaller than $M_P^4$,
or even of the same order of magnitude as  $M_P^4$, if the Universe
is approaching singularity after the corresponding phase transition.
The example of the corresponding plots is shown in Fig.~\ref{brL}. We
can observe that the bounce may hold even for a relatively large
cosmological constant. In fact, the cosmological constant of the given
range does not change, qualitatively, the behavior near the singularity,
even though the phase diagram gets modified. Let us note, without
going into detail, that qualitatively the same situation with bounce
holds for the $k=+1$ space geometry.

\begin{figure}[h!]
\centering
\includegraphics[width=0.42\textwidth]{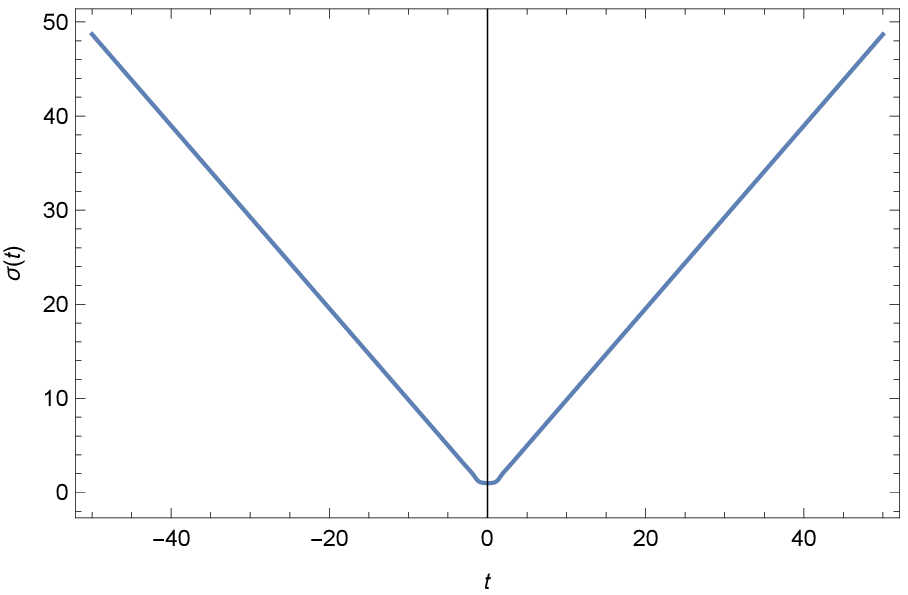}
\includegraphics[width=0.42\textwidth]{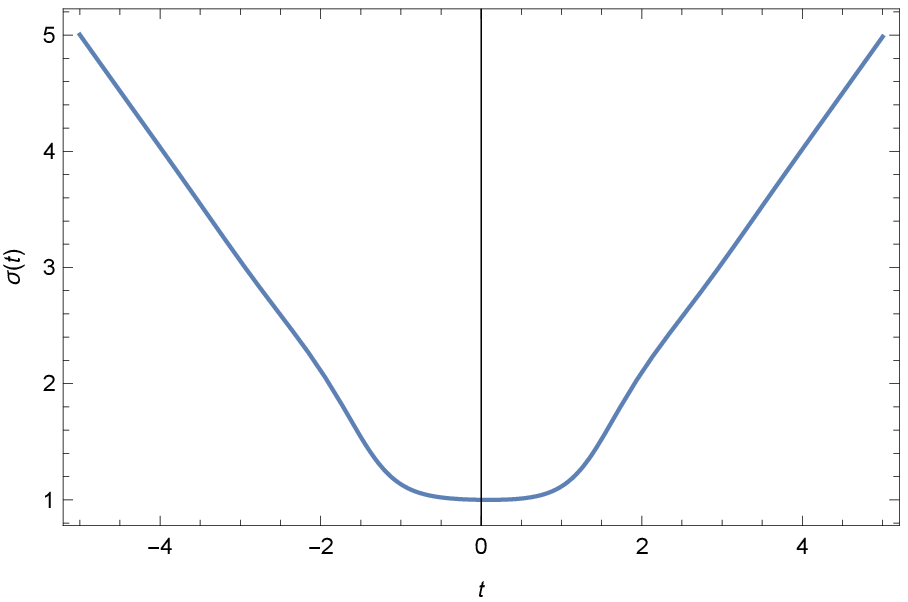}
\begin{quotation}
\caption{\small
Numerical solution for the conformal factor $\si(t)$ in the
presence of the anomalous radiation term. Here, we use the value
$\be \bar{F}^2 = - 0.1$ and the following initial conditions:
$\si(0) = 1$, $\dot{\si}(0)=-0.01H_0$, $\ddot{\sigma} (0) = 0.1$,
and $\dddot {\sigma} (0) = 0 $. In the left plot, we show the
range $-50 \leq t \leq 50$ in Planck units and in the right plot the
range is ten times smaller. One can observed that, even considering
the quantum contribution to radiation, the bounce solution is still
present.}
\label{brd}
\end{quotation}
\end{figure}

The last example of the numerical analysis is included to
illustrate the bounce in the presence of radiation. This is an
important aspect to explore, because in the region close to the
bounce, the solution is rapidly deviating from de Sitter and, as
a consequence, there is an intensive
creation of particles and radiation from the vacuum (see, e.g.,
\cite{Parker1968,ZS-77,DobMar} and references therein).
Thus, there may be a large amount of radiation in the vicinity
of the bounce, and this should be taken into account. However,
since we use the equation for the trace (\ref{T-eq}), the classical
radiation is not ``visible'' for the equation. On the other hand, we
can still model the presence of radiation with the anomalous
contribution, and the result of this is shown in Fig.~\ref{brd}.

\section{Bounce and the stability analysis}
\label{s5}

It is interesting to explore the stability of the bounce solutions
with respect to small perturbations of initial conditions.
The analysis can be easily done analytically using the previous
results (\ref{solapm}) and  (\ref{lapm}) for the exponential
behavior.

According to Eq.~(\ref{H}), there are two pairs of solutions,
corresponding to each of the values (\ref{HH}) with
positive and negative signs. The ``pre-inflationary'' exponential
expansion corresponds to the value of $H_2>0$. Then, from
the diagrams in Figs.~\ref{b2}, \ref{brL}
and \ref{brd},
follows
that the contraction phase corresponds to the change of
sign $H_2 \to -H_2$.

Starting from this point, we can address the issue of stability in
the linear contracting regime $\si(t)= - H_2t$ using the
eigenvalues  (\ref{lapm}) of the solution (\ref{solapm}) for
small perturbations. Since we assume that $a_4$ in
Eq.~(\ref{Einstein}) is zero, the value of $c$ should be defined
from quantum contribution (\ref{c}). It is easy to see that
$1+\frac{8b}{c}<0$ for the MSSM particle contents, when $c>0$.
Then the real parts of both roots $\la_{1/2}$ in Eq.~(\ref{lapm})
are defined by the sign of $H_0$. In the expansion phase,
$H_0=H_2>0$, and both roots have negative real parts. On the
contrary, in case of contraction, we have $H_0=-H_2<0$ and the
solution $\si(t)\approx - H_2t$ is unstable under small
perturbations of $\si(t)$. Thus, the stable bounce due to the
anomaly-induced semiclassical corrections is impossible.

A natural and important question is how it happened that we meet
the bounce solutions as shown in the plots above, if the bounce can
not be stable? As we discussed in the previous section, there is
{\it no} strong dependence on the initial conditions in these
solutions. The numerical analysis shows that the bounce survives
small variations of these conditions, which is typical for the stable
case. We know that typically, for the linearly unstable solutions,
one has to expect that the solution of the desired type can be
achieved only for an absolutely precise fine-tuning of the initial
conditions.

The answer to this question is as follows. Indeed, the stability
or instability depend on the way we interpret the dynamics of the
system or, in other words, depends on our viewpoint.
Defining the initial conditions at the bounce point $t=0$ and
choosing $|\dot{\si}(0)|$ sufficiently small, we are exploring the
contraction phase by looking {\it backward} in time $t$. In this
case, for the contraction there is a negative $t$ and therefore a
positive product $H_0t$. Then Eq.~(\ref{solapm}) tells us that the
contraction phase is stable. However, if we look \ {\it forward} \ in
time, that is choose the initial point at some point $t_0 < 0$ near
the same decreasing line, the solution is unstable.

It is worth noting that the situation in many other bounce models
(see, e.g., the review \cite{Novello2008}) is different. In these
models there is a typical relation $H \sim \frac{1}{t}$, such that
the sign of the product $Ht$ does not change in the transition
between the expansion and contraction phases, when we consider
the dynamics forward in time.\footnote{We are grateful to Nelson
P. Neto for this explanation.}

\section{Conclusions and discussions}
\label{s6}

It is quite natural to expect that the anomaly-induced semiclassical
corrections to the gravitational action cure the cosmological
singularity. Qualitatively, the reason for this expectation is that
the theory with fourth and higher derivatives has a natural cut-off.
In the fourth-derivative theory \cite{Stelle77} this cut-off is the
Planck mass divided by the dimensionless parameter of the relevant
coefficient of the fourth-derivative term. From the dimensional
arguments follows that, in the the cosmological case, the higher
derivative terms should modify the dynamics of the conformal
factor of the metric and remove the singularity.

E.g., in the Newtonian singularity case, i.e., for a point-like
static mass source, many types of higher derivative terms remove
singularity in the modified Newtonian potential. This occurs
starting from the simplest fourth-derivative theory with the action
composed by (\ref{Einstein}) and (\ref{higher}) terms \cite{stelle78}.
The same is true for the local   (see, e.g., \cite{ABS,Breno-PLB}) and
nonlocal \cite{Tseytlin-95} higher derivative models. It is worthwhile
to note that the detailed analysis of the Newtonian and especially
black hole cases is more complicated (see, e.g., \cite{BGTN-18},
\cite{LMBGTN-20} and further references therein).

In cosmology,
the basic four-derivative action (\ref{Einstein}) and (\ref{higher})
does not remove singularity because the $C^2$ term, contributing
with the ghost degree of freedom to the spin-2 sector of the theory,
equals zero on the cosmological background. Then, the $R^2$ term
alone can not help in removing singularity because it does not
produce a massive unphysical ghost.

At this point, it becomes clear that the cosmological case is more
complicated, especially because the situation with singularity can
not be reduced to the analysis of the flat space propagator, as it is
in the Newtonian singularity case. Taking these considerations into
account, one can expect that the anomaly-induced term
(\ref{quantum1}) may remove singularity and promote bounce
since it introduces a more complete form of dynamics to the
conformal factor $\si(t)$. This sort of intuitive arguments should
be carefully verified by direct calculations. In the previous
sections, we saw that the bounce  solutions in this model really
take place. On the other hand, these solutions are unstable if we
define the initial conditions sufficiently far backward in the
contraction regime and look forward in time.
In this respect, the anomaly-induced model is different from the
models with nonlocal form factors, where the bounce solutions
can be found \cite{Biswas2012,Koshelev2013}.

It would be  certainly interesting to obtain a stable bounce
solutions from the semiclassical or quantum gravity corrections,
derived in a consistent way. One of the options is to explore
higher-loop semiclassical corrections in the initially conformal
model, which are expected to give higher than linear powers
of logarithms in the UV form factors and, therefore, stronger
non-linear $\si$-dependence in the effective action (see, e.g.,
\cite{Hamada-2020}). This possibility represents an interesting
proposal for a possible future work.

\section*{Acknowledgments}

Authors is grateful to N.-P. Neto and A.A. Starobinsky for useful
discussions. C.W. is grateful to Coordenação de Aperfei\c{c}oamento
de Pessoal de N\'{\i}vel Superior - CAPES (Brazil) for supporting
this Ph.D. project. The work of I.Sh. was partially supported by Conselho
Nacional de Desenvolvimento Cient\'{i}fico e Tecnol\'{o}gico - CNPq
(Brazil) under the grant 303635/2018-5.
I.Sh. was also supported in parts by Russian Ministry
of Science and High Education, project No. FEWF-2020-0003.


\end{document}